\documentclass[letterpaper]{article}

\usepackage{natbib,alifeconf}  

\usepackage{mdframed}
\usepackage{hyperref}
\usepackage{multirow}
\usepackage{changepage}

\title{\vspace{-2cm}Cities as they could be: Artificial Life and Urban Systems}
\author{Juste Raimbault$^{1,2,3}$\\
\mbox{}\\
$^1$Center for Advanced Spatial Analysis, University College London\\
$^2$UPS CNRS 3611 ISC-PIF\\
$^3$UMR CNRS 8504 G{\'e}ographie-cit{\'e}s\\
juste.raimbault@polytechnique.edu} 

\begin{document}
\maketitle

\begin{abstract}
The metaphor of cities as organisms has a long history in urban planning, and a few urban modeling approaches have explicitly been linked to Artificial Life. We propose in that paper to explore the extent of Artificial Life and Artificial Intelligence application to urban issues, by constructing and exploring a citation network of around 225,000 papers. It shows that most of the literature is indeed application of methodologies and a rather strong modularity of approaches. We finally develop ALife concepts which have a strong potential for the development of new urban theories.
\end{abstract}


\section{Introduction}

The understanding of processes driving the growth of cities, or more generally the evolution of urban systems, and approaches for a sustainable design and management of these, have for more than 100 years been tightly linked to concepts borrowed from biology such as evolution \citep{batty2009centenary}. Inventing novel ways to design cities, beyond predicting their evolution \citep{batty2018inventing}, is a major asset to tackle climate change and sustainability issues \citep{iepsfc2018science}. In that context, Artificial Life (ALife) approaches to urban systems have several advantages, including knowledge transfer from biology and ecology where relevant concepts such as resilience or morphogenesis are thoroughly studied, a strong practice of interdisciplinarity, or methodologies and tools such as agent-based modeling and cellular automata, among others. Following the seminal view of ``Life as it could be'' by \cite{langton1986studying}, an explicit ALife take on urban issues would consist in studying ``Cities as they could be''. Although not explicitly listed in application domains of ALife by \cite{kim2006comprehensive}, is is included through the use of methods and in relation with economic models. Two of archetypal ``would-be'' worlds of \cite{casti1997would} are related to urban systems (transportation and resource exploitation). \cite{doi:10.1162/isala00135} recently introduced a conceptual frame considering cities from an organismic perspective, while \cite{doi:10.1162/isala00134} showed that open-endedness has a strong potential to develop sustainable social systems.

Diverse streams of research on urban systems have already linked with ALife. From a methodological viewpoint, the use of cellular automata and agent-based models for urban growth and urban dynamics has a long history \cite{torrens2003automata}. These can be used for example to generate building layouts and road networks of synthetic cities \citep{kato1998alife} or at the scale of districts \citep{raimbault2019generating}. Integrating such models into evolutionary computation algorithms widens the scope of possible synthetic cities \citep{kato2000modeling}. Other dimensions such as land prices and residential dynamics can be grasped using agent-based models \citep{takizawa2000simulation}. Generative processes can also be used for interactive urban design \citep{openshaw1995developing}. The study of urban morphology may be done with methods used to study morphogenesis, and \cite{medda2009morphogenetic} apply reaction-diffusion equation to model the relation between transportation and land-use, while \cite{raimbault2018calibration} shows that the combination of aggregation and diffusion produces realistic urban forms. \cite{d2013simulating} explores possible future sustainable urban morphologies. \cite{lucic2002transportation} use bio-inspired algorithms to solve difficult transportation planning problems. The concept of urban metabolism introduced by \cite{olsen1982urban} also comes as a transfer from biology.

Moreover, the field of Artificial Intelligence (AI) has also numerous application related to urban systems. \cite{wu2010artificial} review AI application to the prediction of land-use dynamics, unveiling a very broad range of methods ranging from evolutionary computation to neural networks, and suggesting that integrative and interdisciplinary approaches still lack for more robust urban applications. \cite{white1989artificial} show for example that a neural network trained appropriately can learn to plan transportation infrastructures. \cite{zheng2014urban} define the emerging field of Urban Computing as the convergence of ubiquitous urban data with artificial intelligence and new urban services, with varied domains of application including transportation, economy, environment, and planning. AI can for example be applied in real-time conditions to manage highway traffic \citep{ma2009real}. Other urban dimensions which require accurate predictions with a high spatio-temporal resolution, such as water demand \citep{adamowski2010comparison} or electric vehicles grid management \citep{rigas2014managing}, are other examples where AI is successfully applied.

Thus, as both ALife and AI have been broadly applied to urban systems, we can first ask what are their respective extent in terms of methodologies, tools, concepts, and application domains, and secondly what are their remaining potentialities to enhance the understanding and management of cities, in other words what research directions and concepts in that particular context remain to be explored. This paper proposes to tackle these two questions by means of a systematic literature mapping method based on citation networks. More precisely, our contribution (i) constructs and explores a large citation network of around 225,000 papers, to map the respective contributions of ALife and AI to the urban literature, and their relations; and (ii) explores more thoroughly crucial concepts still loosely applied or understood in an urban system perspective.

The rest of this paper is organized as follows: we develop in the next section the bibliometric analysis based on citation networks to map the scope of artificial life and artificial intelligence approaches to urban systems. Building on this systematic mapping, we then review principal points in which artificial life can significantly inform the study of urban systems, which can be understood as research directions opened by taking such a viewpoint of ``Cities as They Could Be''.

\section{Bibliometric analysis}



Literature mapping and quantitative bibliometrics have been widely used to reinforce knowledge in most disciplines, and are part of a field of study in itself \cite{leydesdorff2001challenge}. They are furthermore important to enhance reflexivity which is crucial in disciplines studying socio-technical systems \cite{raimbault2019empowering}. In the case of artificial intelligence, several mappings have been proposed, for example from a semantic \citep{van1993neural}, spatialized \citep{niu2016global}, or journal-level \citep{ibanez2011using} viewpoint. \cite{squazzoni2013social} analyze the impact of the Journal on Artificial Societies and Social Simulation. \cite{aguilar2014past} show the evolution of theme frequency in time for the Artificial Life journal. There is however to the best of our knowledge no previous attempt of such an exercise for Artificial Life at a large scale. We propose here such a literature mapping approach to both ALife and AI, in the specific context of urban systems applications.

We use therefore a citation network analysis, applying the methods and tools developed by \cite{raimbault2019exploration}. In a nutshell, citation networks are constructed by first constituting a seed corpus using a keyword search, and then by collecting papers citing papers in this corpus, recursively to a certain level. This method is an effective tool to reconstruct endogenous disciplines of citation practice. It is implemented into an open source library collecting and processing data from Google scholar, available at \url{https://github.com/JusteRaimbault/BiblioData}.


The keyword we consider for constructing the initial corpus are \texttt{Artificial Intelligence AND Urban} for AI approaches and \texttt{Artificial Life AND Urban} for ALife. Using alternative terms such as ``machine learning'' or ``city'' expands the scope too much and yield non relevant results. A size of 200 papers for each request appears to cover most of relevant literature, the higher rank query results being more and more unrelated. The citation propagation will in any case retrieve possibly missing papers. 25 references in the seed corpus which obviously did not fit the scope were manually removed (sociology or gender studies e.g.), what yields a corpus of 296 papers (due to the overlap between the two requests). From this layer, the backward citation network is retrieved up to depth two. This ensures in particular that all links within the depth one layer are effectively observed.


Data collection and preprocessing is done in Java using the BiblioData library, and network analysis is done in R with the igraph package. Network visualization is done with the Gephi software. All raw and processed data is available on dataverse at \url{https://doi.org/10.7910/DVN/RXBJ7R}, and source code and results are available on a git repository at \url{https://github.com/JusteRaimbault/UrbanEvolution}.





The raw citation network contains $\left|V\right| = 224,510$ papers and $\left|E\right| = 315,829$ citation links. We keep the largest connected component (size $223,476$ against the second of size only $664$), and furthermore extract the core of the network by removing all nodes of degree 1, and repeating this until all nodes have at least a degree of two. We then work with this core network with $48,657$ nodes and $139,931$ edges.






\begin{table*}[h!]
\begin{adjustwidth}{-1cm}{-1cm}
\center{
\begin{tabular}{|c|c|c|c|c|c|}\hline
\multicolumn{3}{|c|}{Full network} & \multicolumn{3}{|c|}{ALife subnetwork}\\\hline
Community & Size (\%) & Repr. paper & Community & Size (\%) & Repr. paper\\ \hline\hline
Real-time mapping & 13.5 & \citep{thrun2006graph} & Complexity & 12.7 & \citep{casti1997would} \\
Autonomous vehicles & 9.1  & \citep{levinson2011towards} & Flying UAV & 11.0 & \citep{kendoul2012survey} \\
Traffic forecasting & 7.1  & \citep{vlahogianni2005optimized} & Cellular automata & 9.8 & \citep{white1993cellular} \\
3D Modeling & 6.5  & \citep{akbarzadeh2006towards} & Autonomous vehicles & 9.1 & \citep{gonzalez2015review} \\
Complexity/ABM & 5.3  & \citep{casti1997would} & Culture & 8.6 & \citep{franklin2002embodied} \\
Computer vision & 5.0  & \citep{sen2004robust} & Crowd behavior & 7.2 & \citep{pelechano2007controlling} \\
Traffic control & 4.9  & \citep{chen2010review} & Traffic control & 6.5 & \citep{chen2010review} \\
Urban growth/CA & 4.9  & \citep{batty1994cells} & Traffic forecasting & 4.5 & \citep{lv2014traffic} \\
Rescue robots & 4.8 & \citep{davids2002urban} & Social intelligence & 4.1 & \citep{wang2007toward} \\
Art/Culture & 4.3 & \citep{whitelaw2004metacreation} & Urbanism & 4.0 & \citep{schipperijn2010factors} \\\hline
\end{tabular}
}
\end{adjustwidth}
\caption{Largest citation communities, in the full core network ($\left|V\right| = 48,657$) and in the ALife-induced core network ($\left|V\right|=17,705$). We give the relative share of each community and one representative paper, among the ones with the largest degree.\label{tab:coms}}
\end{table*}

We use the Louvain algorithm for community detection \citep{blondel2008fast} on the undirected equivalent network to reconstruct endogenous scientific fields. We obtain 33 communities, with a directed modularity \citep{nicosia2009extending} of 0.84, what corresponds to highly clustered communities. Their sizes are relatively balanced, since the size distribution with a cutoff of 500 papers has a Zipf exponent (OLS estimation) of 0.73. We associate a ``field'' to each community by inspecting the largest degree papers in each. We describe the ten largest community in Table~\ref{tab:coms} with their label, relative size, and representative paper. The content is broad in terms of disciplines involved, methods used, and applications domains. More precisely, they can be grouped into the following main themes:

\begin{itemize}
	\item Autonomous vehicles, i.e technologies developed for self-driving vehicles in urban environments (Autonomous vehicles, Real-time mapping, Flying UAV, Image capture, Power line detection, Lidar mapping, Computer vision);
	\item Robots: use rescue robots in urban disasters;
	\item Environmental science: application of mostly AI methods to the estimation of environmental variables or natural hazards (Natural hazards, Flooding, Rainfall, Air pollution, Biology);
	\item Traffic and transportation, in terms of management and planning (Traffic control, Traffic forecasting, Traffic safety, Transit planning, Bio-inspired computing applied to transit);
	\item Urban growth: analysis and modeling of urban growth (Urban growth/Cellular automata, Remote sensing, GIS);
	\item Social systems: simulation of artificial societies and agent-based modeling (Social intelligence, Game theory, Fuzzy cognitive maps, Crowd simulation, Smart cities);
	\item Complexity: concepts and methods linked to complexity, mostly from ALife (Complexity/Agent-based Modeling, Self-organization)
	\item Architecture: bio-inspired and computational architecture and urbanism (Living architecture, Urbanism, 3D Modeling);
	\item Humanities: art and cultural studies such as generative art and studies of digital culture (Art/Culture). The emergence of this community is interesting in itself, as it shows the importance of artistic approaches in the ALife landscape, actually witnessed in recent ALife conferences with the Art track, and more generally by the crucial role of epistemology and philosophy in complexity science.
\end{itemize}

\begin{figure*}
\begin{center}
\begin{mdframed}
\includegraphics[width=\linewidth]{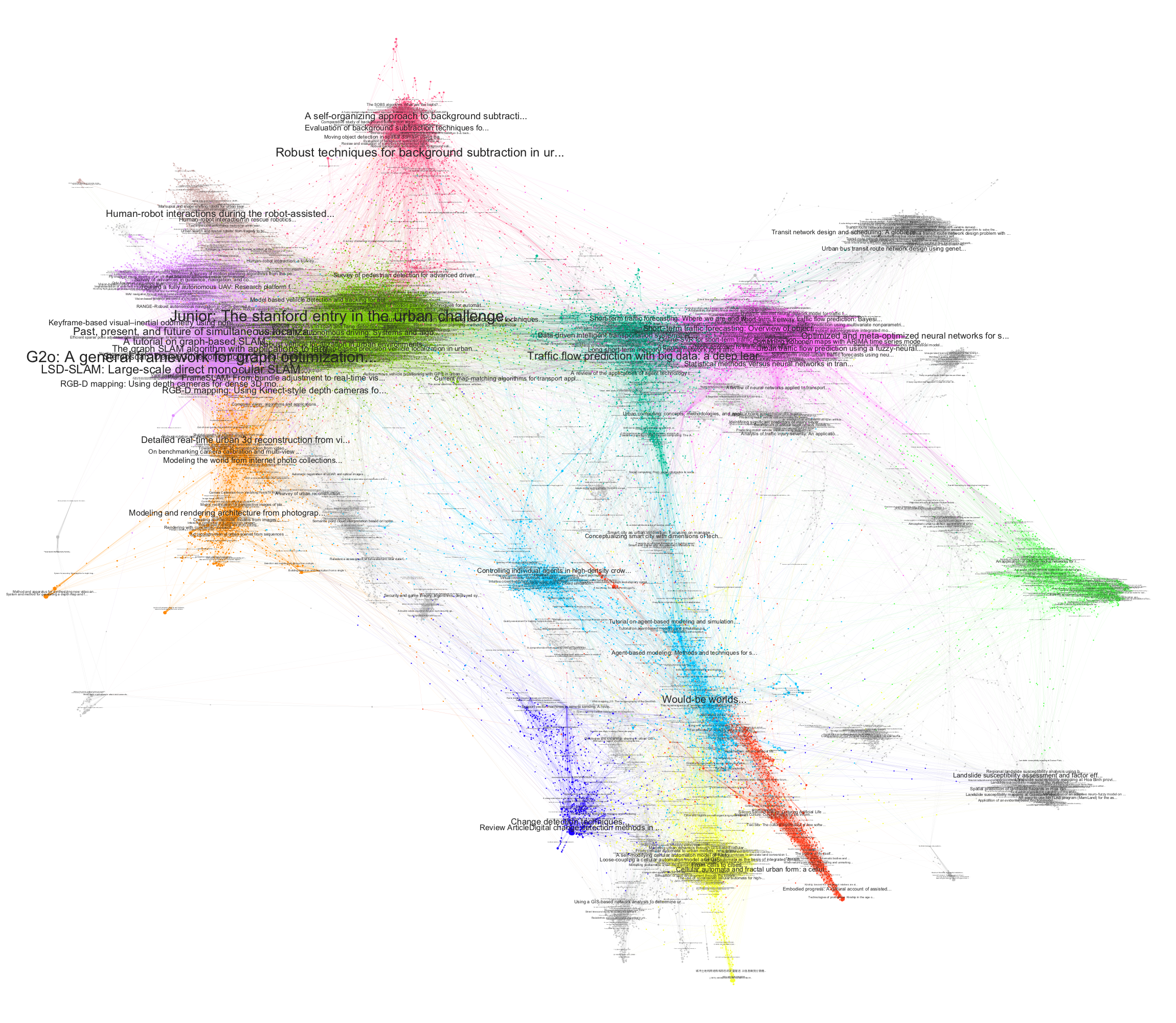}
\end{mdframed}
\begin{mdframed}
\includegraphics[width=\linewidth]{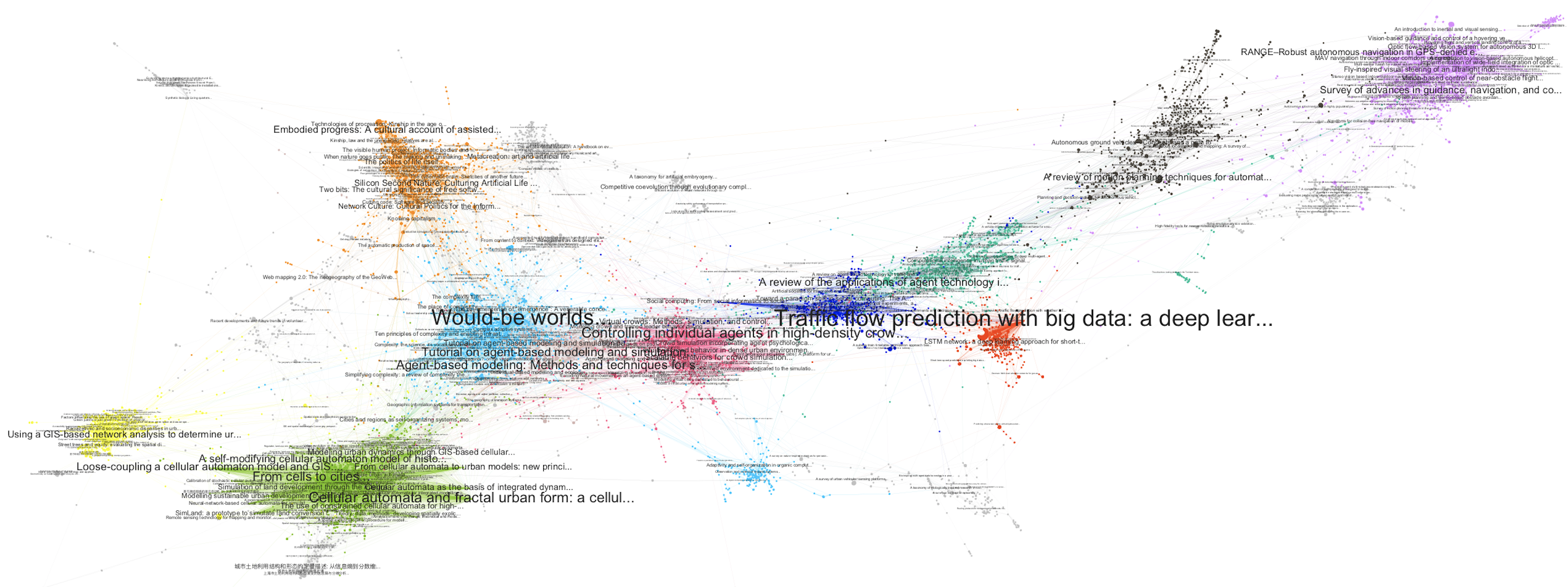}
\end{mdframed}
\caption{Citation networks for Artificial Life and Artificial Intelligence applications to urban systems. Full core citation network (top panel) and ALife subnetwork (bottom panel). Network visualization is done with the Gephi software, spatialization with a Force Atlas 2 algorithm, node label size corresponds to node degree. Colors correspond to communities obtained with the Louvain algorithm.\label{fig:citnw}}
\end{center}
\end{figure*}

Note that these main themes do not necessarily correspond to the actual organization of communities and their interactions. To grasp these in a first order simple manner, we can inspect the network visualization shown in Fig.~\ref{fig:citnw} (top panel). In the network spatialization, we observe a big cluster of communities formed by studies on autonomous vehicles and robotics (top left of the network). As real-time mapping for such technologies is closely related to 3D reconstruction used in architecture, this last community is far from other architecture communities which are closer to ALife approaches on the bottom. An other large and relatively isolated cluster corresponds to the transportation theme (top right cluster), while more sparse communities, mostly related to ALife (complexity, urban growth, artificial societies, art and culture), gather on the bottom of the network. These include Urban growth modeling, Remote sensing and GIS, but also the complexity and Art/Culture community. Finally, Environmental science communities, such as pollution or landslides, are mostly quite isolated (e.g. the two on the most right) as they are highly specific to the application. The ``bridge'' between ALife approaches on the bottom of the network and the approaches more dominated by AI on the top (transportation and autonomous vehicles) is actually done by communities related to social systems: Crowd simulation, Smart cities and game theories. In a sense, ALife focuses more on generative processes for the built environment, while AI on technical issues related to urban technologies, what makes the ``human-in-the-middle'' a logical mediator between the two. Indeed, urban systems are complex socio-technical systems, in which human agents interact between themselves, and multiple social and technical layers.

\begin{figure*}
\vspace{-1cm}
\begin{center}
\includegraphics[width=0.49\linewidth]{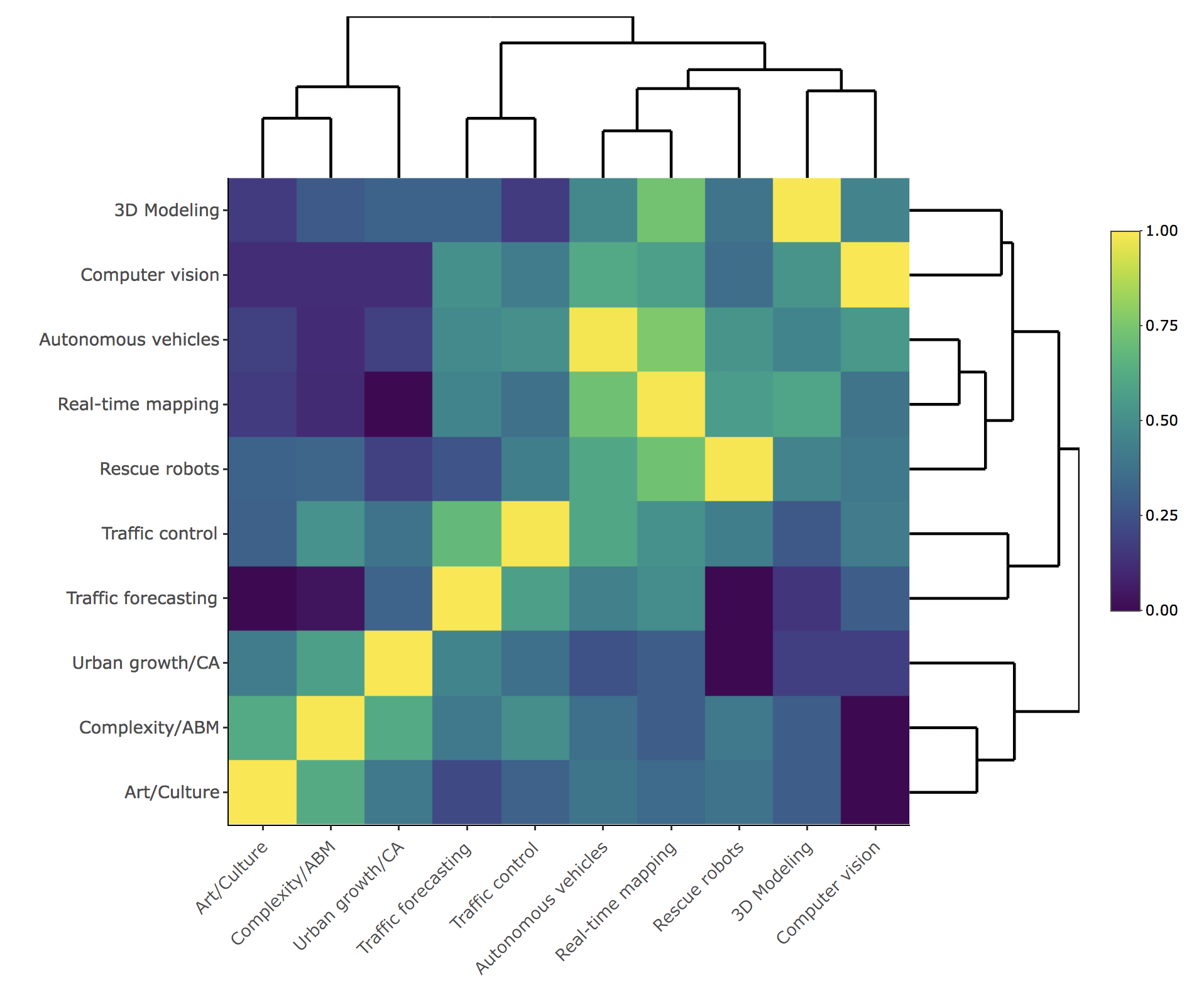}
\includegraphics[width=0.49\linewidth]{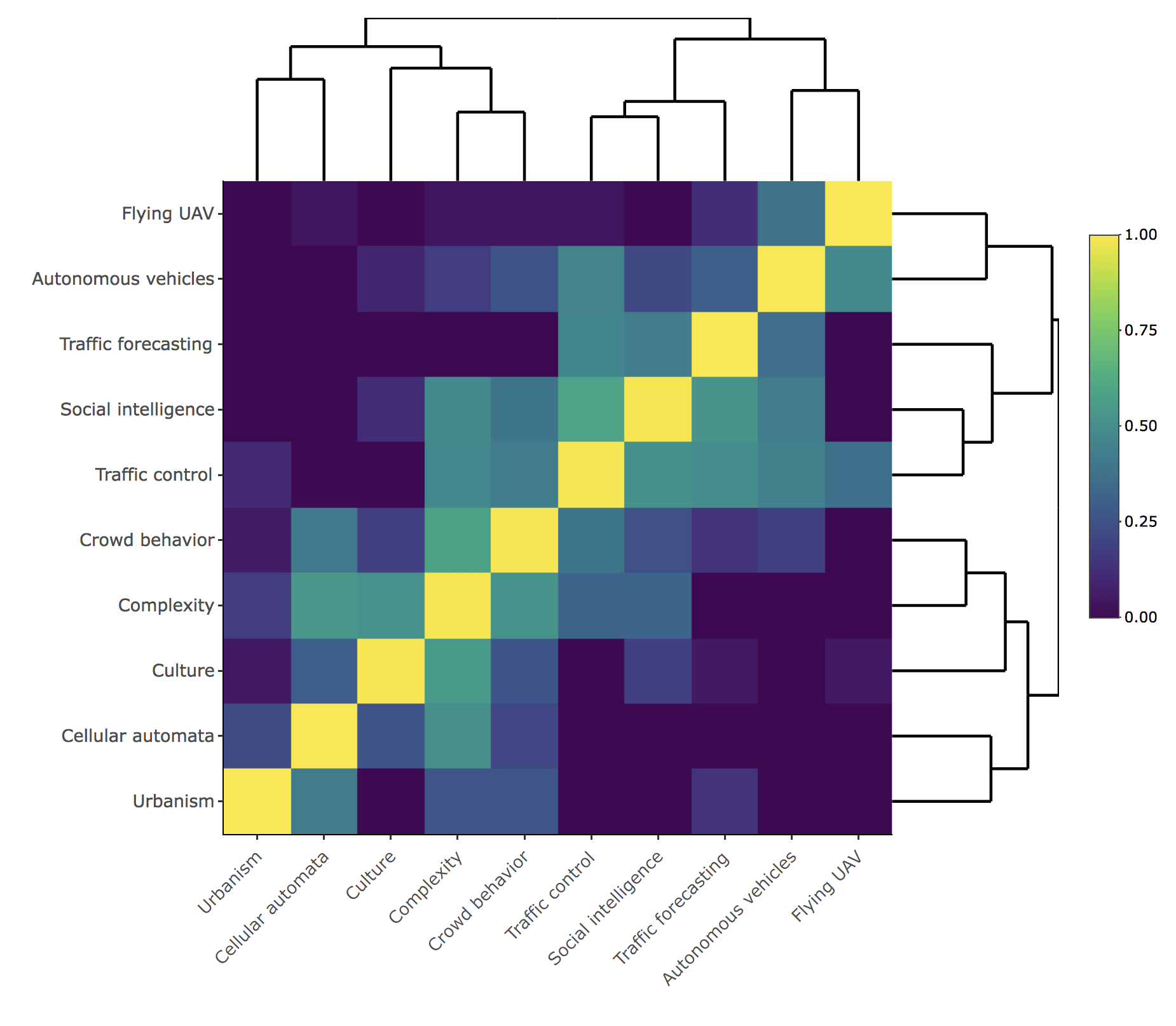}
\caption{Citation flows between largest communities, for the full network (Left) and the ALife subnetwork (Right). Scale color is given by a renormalized logarithmic scale (the major share of citations being internal to communities), the maximal flow being 0.95 (resp. 0.98) for the full network (resp. the ALife subnetwork) and the minimal flow $4.8\cdot 10^{-5}$ (resp. $1.9 \cdot 10^{-4}$). Dendograms show a hierarchical clustering following the proximity matrix.\label{fig:flows}}
\end{center}
\end{figure*}

To confirm which are the communities typical to ALife but not AI approaches, we study the subnetwork originating from the ALife keyword request only. In terms of size, the core of this network has only $\left|V\right|=17,705$ papers, which is a bit more than a third of the full network. Artificial Intelligence is thus larger than ALife for paper quantity in urban applications, what is consistent with the respective size of their literature and communities. We also give the 10 largest ALife citation communities in Table~\ref{tab:coms}, and find quite a high overlap in term of content. Not surprisingly, the largest is the conceptual community related to complexity and agent-based modeling, dominated by \cite{casti1997would} which surveyed two typical artificial societies. Autonomous vehicles and robotics are still present, what is consistent with the ``hard'' branch of ALife linked to self-assembly and self-organizing robots. When inspecting the spatialized subnetwork (Fig.~\ref{fig:citnw}, bottom panel), we find that the Complexity community is central, and that social modeling and transportation are this time making the bridge with autonomous vehicles, which are weakly linked to the rest of the network. The Remote Sensing community has disappeared, what is consistent with the fact that this aspect of urban studies corresponds mostly to machine learning for unsupervised learning. Similarly, Smart cities do not appear in the ALife subnetwork: indeed, most of these approaches do not deal with slow time scales nor with generative processes, being in a sense ``superficial'' since they do not capture underlying processes governing the evolution of urban systems, but aim at exploiting a new technical layer (consisting of sensors e.g.) put on top of existing urban structures.






Finally, we can quantify the relation between the citation communities in terms of citation flows. More formally, if $c_{ij}$ is the citation network adjacency matrix, the directed flow from community $k_1$ to community $k_2$ is the share of citation made by $k_1$ which go to the other community, what reads $\phi_{i \rightarrow j} = \frac{\sum_{i\in k_1,j\in k_2} c_{ij}}{\sum_{i\in k_1,j} c_{ij}}$. These flows are shown as heatmaps in Fig.~\ref{fig:flows}. We observe consistent clustering of communities, following previous results from network spatialization and main themes: for example, Autonomous vehicles, 3D Modeling, Computer vision and Real-time mapping, form a strong cluster in the full network, as traffic control and traffic forecasting do. Note that flows are not symmetric: for example, Art/culture cite a small proportion of literature related to traffic, but traffic is totally blind to it. A similar asymmetry is observed between rescue robots citing traffic without reciprocity. We find that the ALife subnetwork is more modular, with less flows between far communities, but still a progressive shift from humanities (Urbanism) to technical disciplines (Autonomous vehicles), with successive sub-blocks on the diagonal corresponding to clusters of communities. In a nutshell, we confirm the relations unveiled previously, and that domains remain isolated in absolute terms (we recall that heatmap have a logarithmic scale).

From this systematic literature mapping and exploration, we can conclude the following main stylized facts: (i) AI and ALife have overlapping applications for urban systems, but still remain complementary altogether; (ii) the contributions are mostly methodological and technical, since the only community explicitly including conceptual contributions (Complexity) represents only 5.3\% of the network. Similarly, approaches dealing with generative urban processes on slow time scales, typically urban morphogenesis, which dominated our introducing literature review which was driven by ALife concepts, is a relatively restreint part of the full corpus (around 10\% by including CA and ABM).




\section{Discussion: Cities as they could be, strengthening ALife concepts in urban science}

We turn now to a more thorough development of some concepts related to ALife which should have either a high theoretical importance for the study of urban systems, or a high potentiality to introduce novel approaches. This list stems from the conjunction of open issues in urban science \citep{lobo2020urban} with conclusions from the previous literature mapping on underexplored paths. It can be understood as main arguments of why ALife concepts may help understanding, planning, designing and managing in a better and more sustainable way.



\subsection{Tools and methods}


We saw that tools and methods, in particular Artificial Intelligence methods, are already widely applied. However, new methods from Artificial Life, such as modeling frameworks which would be appropriate to simulate open-ended evolution \citep{banzhaf2016defining}, could have an important role in exploring potential urban futures by means of simulation.

\subsection{Biomimicry and urban ecology}


\cite{taylor2017art} has put forward biomimicry as a powerful approach for an innovative urban design, indeed in a way similar to evolutionary computation and bio-inspired computing or engineering following natural design principles. Building on the broad and old literature on metaphors of the city as an organism, these approaches remain mostly to be explored and tested, preferentially using simulation models, and transferring existing knowledge from ALife for this. Urban metabolism enters this frame, and \cite{raimbault2020spatial} for example systematically explore an agent-based model for industrial symbiosis.

\subsection{Morphogenesis and autopoiesis}


The link between urban form and urban function remains an open fundamental question in the study of urban systems. The field of morphogenetic engineering provides a framework to engineer this relation from the bottom-up \citep{doursat2012morphogenetic}, which fits particularly well issues in urban planning and design. How it can be concretely applied is however still unknown. Furthermore, beyond the concept of morphogenesis and urban morphogenesis models, the concept of autopoiesis could play a crucial role in the study of urban systems. An autopoietic system can be understood as a network of processes with well-defined boundaries and which self-sustains itself \citep{bourgine2004autopoiesis}. The issue of urban systems boundaries is crucial in the definition of urban systems (see e.g. \citep{cottineau2017diverse} in the case of scaling laws), and investigating the autopoietic nature of cities should be a promising research direction.

\subsection{Urban evolutionary theories}


The definition of an ``Urban DNA'', i.e. an extension of the gene concept in evolution or of the \emph{meme} concept in cultural evolution, or even an approach combining different types of replicators \citep{bull2000meme}, remains an open question. Previous attempts such as \citep{votsis2019urban} are not satisfying since they rely more on the extraction of phenotypical parameters based on urban form. Fundamental processes to effectively have evolution (such as transmission, exchanges and mutation) work well in the case of cultural evolution, but urban systems have also slow time scale artefacts and infrastructures which may follow other types of rules. Defining and understanding co-evolution processes in urban systems is also an open research avenue strongly related to ALife. In that context, \cite{raimbault2019modeling} proposed for example a co-evolution approach to the particular case of modeling interactions between transportation networks and territories.

\subsection{Innovation in urban systems}


In relation to the previous point, \cite{pumain2018evolutionary} proposes an urban evolutionary theory based on the diffusion of innovation between cities. It is evolutionary in the sense that cities are complex adaptive systems of systems. Defining innovation and characterizing it in artificial systems is one cornerstone of ALife research, in particular through studies of open-ended evolution. In that sense, as innovation is indeed central to urban systems (technological, but also social and cultural), a better understanding of it is needed.

\subsection{Linking Urban ALife and AI: urban computing}


Finally, a last point possibly joining AI and ALife potentialities for urban systems, is the concept of urban computing. As currently defined, it relates more to AI approaches. However, in the same way that biological systems can compute, the notion of social and collective computation \citep{flack2011challenges} relates to social intelligence and collective decision-making. To what extent are these processes modified when embedded into territorial systems, in particular through the geographical constraints, and to what extent new hybrid systems integrating sensors and crowd-sourced data can actually compute and make cities ``smarter'', i.e. in a sense how to achieve territorial intelligence, has still to be investigated.

%

\section{Conclusion}

We have developed a literature mapping of AI and ALife approaches to urban systems. It appears that most application remain methodological and technical, and that many conceptual research avenues stemming from Artificial Life remain to be developed as new theories of urban systems. We believe these have a strong potential for sustainable urban planning and management.


\footnotesize

\end{document}